\documentclass[useAMS,usenatbib]{mn2e}
\usepackage{rotating}
\usepackage{longtable}
\usepackage{graphics}
\usepackage{times}

\def\teff{$T_{\rm eff}$}
\def\cvn{$\alpha^2$\,CVn}
\def\lgg{$\log g$}

\def\kms{km\,s$^{-1}$}
\def\i{\,{\sc i}}

\def\fe{Fe\,{\sc ii}}
\newcommand{\lgfe}[1]{$\log N_{\rm Fe}/N_{\rm tot}=#1$}

\newcommand{\inva}{\mbox{\sc Invers10}}
\newcommand{\invb}{\mbox{\sc Invers13}}
\newcommand{\llm}{\mbox{\sc LLmodels}}

\def\aj{AJ}%
\def\apj{ApJ}%
\def\apss{Ap\&SS}%
\def\aap{A\&A}%
\def\aaps{A\&AS}%
\def\mnras{MNRAS}%
\def\pasp{PASP}%
\newcommand{\figps}[1]{\resizebox{\hsize}{!}{\rotatebox{0}{\includegraphics{#1}}}}

\newcommand{\fifps}[2]{\centering\resizebox{#1}{!}{\includegraphics{#2}}}

\begin{document}

\title[Magnetic Doppler imaging considering metallicity-dependent atmospheric structure]
{Magnetic Doppler imaging considering atmospheric structure modifications due to local abundances: a luxury or a necessity?}

\author[O.~Kochukhov, G.~A.~Wade and D.~Shulyak]
{O.~Kochukhov$^1$, G.~A.~Wade$^2$ and D.~Shulyak$^3$ \\
$^1$ Department of Physics and Astronomy, Uppsala University Box 516, 751 20 Uppsala, Sweden\\
$^2$ Department of Physics, Royal Military College of Canada, Box 17000, Stn Forces, Kingston, Ontario K7K 7B4, Canada\\
$^3$ Institute of Astrophysics, Georg-August University, Friedrich-Hund-Platz 1, D-37077 G\"ottingen, Germany}

\date{Accepted 2012 January 9.  Received 2012 January 9; in original form 2011 December 6} 

\pubyear{2011} 

\maketitle

\label{firstpage}

\begin{abstract}
Magnetic Doppler imaging is currently the most powerful method of interpreting high-resolution spectropolarimetric observations of stars. This technique has provided the very first maps of stellar magnetic field topologies reconstructed from timeseries of full Stokes vector spectra, revealing the presence of small-scale magnetic fields on the surfaces of Ap stars. These studies were recently criticisied by Stift et al., who claimed that magnetic inversions are not robust and are seriously undermined by neglecting a feedback on the Stokes line profiles from the local atmospheric structure in the regions of enhanced metal abundance. We show that Stift et al. misinterpreted published magnetic Doppler imaging results and consistently neglected some of the most fundamental principles behind magnetic mapping. Using state-of-the-art opacity sampling model atmosphere and polarised radiative transfer codes, we demonstrate that the variation of atmospheric structure across the surface of a star with chemical spots affects the local continuum intensity but is negligible for the normalised local Stokes profiles except for the rare situation of a very strong line in an extremely Fe-rich atmosphere. For the disk-integrated spectra of an Ap star with extreme abundance variations, we find that the assumption of a mean model atmosphere leads to moderate errors in Stokes $I$ but is negligible for the circular and linear polarisation spectra. Employing a new magnetic inversion code, which incorporates the horizontal variation of atmospheric structure induced by chemical spots, we reconstructed new maps of magnetic field and Fe abundance for the bright Ap star \cvn. The resulting distribution of chemical spots changes insignificantly compared to the previous modelling based on a single model atmosphere, while the magnetic field geometry does not change at all. This shows that the assertions by Stift et al. are exaggerated as a consequence of unreasonable assumptions and extrapolations, as well as methodological flaws and inconsistencies of their analysis. Our discussion proves that published magnetic inversions based on a mean stellar atmosphere are highly robust and reliable, and that the presence of small-scale magnetic field structures on the surfaces of Ap stars is indeed real. Incorporating horizontal variations of atmospheric structure in Doppler imaging can marginally improve reconstruction of abundance distributions for stars showing very large iron overabundances. But this costly technique is  unnecessary for magnetic mapping with high-resolution polarisation spectra.
\end{abstract}

\begin{keywords}
stars: chemically peculiar --
stars: atmospheres --
stars: magnetic fields --
stars: spots --
stars: individual: $\alpha^2$\,CVn
\end{keywords}

\section{Introduction}

Doppler Imaging (DI) of stellar surfaces represents one of the spectacular successes of modern observational stellar astrophysics. Since the first applications of these tomographic modeling techniques to cool and hot stars in the 1970s and early 1980s \citep{khokhlova:1975,goncharskii:1977,goncharskii:1982,vogt:1983a}, they have revealed the complex surface structures of rotating stars in remarkable detail, providing important new insights into the processes occurring in stellar outer layers and unique constraints on theoretical models \citep{donati:2003,donati:2006,donati:2006b,petit:2008a,morin:2008,kochukhov:2007b,kochukhov:2010}.

Conventional DI interprets rotational modulation of the intensity spectra of stars to reconstruct two dimensional maps of photospheric temperature or abundance spots (in the case of cool and hot stars, respectively). Basically, the stellar surface is divided into a grid of pixels, each of which is assigned independent local physical properties (temperature or abundance). Synthetic disc-integrated spectra are computed, taking into account the local temperature/abundance of each pixel, as well as its rotational Doppler shift, and compared to a timeseries of observed spectral line profiles. Iterative adjustment of the local physical conditions leads to a model (or ``map'') that is capable of reproducing the observed line profile shapes and modulation in detail.

As the available observational data and computational resources have improved, DI models have become increasingly sophisticated, with better resolution and realism (e.g. inclusion of various physical effects such as differential rotation). A qualitative step forward was taken with the inclusion of magnetic fields by \citet{piskunov:1983} for Ap stars and by \citet{semel:1989} for cool active stars. Zeeman or Magnetic Doppler Imaging \citep[ZDI or MDI;][]{brown:1991,piskunov:2002a,kochukhov:2002c} extends the basic principles of DI to allow mapping of stellar magnetic fields by interpreting timeseries of line profile Zeeman polarisation. Currently, the most sophisticated MDI codes \citep{piskunov:2002a,kochukhov:2002c} are capable of simultaneous and self-consistent mapping of surface abundances or temperature distributions along with the vector magnetic field from timeseries of line profiles in two ($IV$) or four ($IQUV$) Stokes parameters.

Recently, \citet*{stift:2011} reported the results of experiments that seriously question the basic reliability of the results of DI and MDI, particular as applied to chemically peculiar Ap stars. As pointed out by Stift et al., DI techniques have revealed some surprising and exotic characteristics of the atmospheres of these objects. One of the most outstanding is the large chemical abundance contrasts in atmospheres of Ap stars, which according to DI analyses can be quite extreme (e.g. variations of several orders of magnitude across the stellar surface). In regions of high enrichment, some metals are sometimes inferred to be only a factor of $\sim$\,30 less abundant compared to hydrogen \citep[e.g.,][]{kuschnig:1998a}, which is challenging to explain theoretically. The distributions of abundances can be extremely complex \citep{kochukhov:2004e}, while sometimes they are rather simple \citep[e.g.,][]{luftinger:2010}. The distributions may show straightforward relationships with the inferred magnetic field geometry \citep*[e.g.,][]{rice:1997}, but frequently they do not. Recently, MDI mapping using four Stokes parameter datasets \citep{kochukhov:2004d,kochukhov:2010} has revealed the unexpected presence of intense, small-scale structures in the reconstructed magnetic field. The existence of these structures is unexplained, and adds a qualitatively new element to our growing picture of the magnetic structure of Ap stars.

The essential claim of \citet[][hereafter S12]{stift:2011} is that maps computed for Ap stars using DI and MDI methods are fundamentally unreliable and that the variety of unexplained properties of maps described above are artefacts of neglecting the impact of large abundances and abundance contrasts on the local stellar atmospheric structure \citep[e.g.,][]{khan:2007}. In particular, Stift et al. state that ignoring these effects results in artificial line profile variability, which translates into spurious abundance and magnetic structures in DI maps. Ultimately, they conclude that complex DI maps with high-contrast abundance distributions are incorrect, and that the presence of small-scale structure and high-contrast magnetic spots, as inferred from MDI, cannot at present be reliably ascertained, and are likely an artefact of the inversion procedure.

In this paper we confront the assumptions and methodology of S12, demonstrating that their conclusions are largely founded on fundamental errors and unrealistic assumptions, and concluding that DI maps based on single, mean stellar atmospheres are highly robust and reliable. Our paper is organised as follows. First, in Section~\ref{general} we bring to the reader's attention several key aspects of DI and MDI methodology overlooked or misinterpreted by S12. Calculation of realistic opacity sampling model atmospheres, corresponding local continuum intensities, and the Stokes $IQUV$ profiles is described in Section~\ref{local}. Using the Fe abundance distribution and magnetic field topology reconstructed for \cvn\ by KW10, we compute in Section~\ref{global} the full Stokes profiles of a strong \fe\ line, exploring the consequences of the lateral variation of atmospheric structure on the intensity and polarisation profiles. A complete MDI inversion based on a grid of model atmospheres with different elemental abundances is presented for \cvn\ in Section~\ref{inversion}. We summarise and discuss results of our investigation in Section~\ref{discussion}.

\section{General considerations}
\label{general}

\subsection{Robustness and stability of magnetic inversions}

As with any complex data modelling technique, capabilities and intrinsic limitations of MDI must be tested comprehensively. The physical foundations and numerical methods implemented in the \inva\ code were described in detail by \citet[][hereafter PK02]{piskunov:2002a}. This paper presented a general discussion of the full Stokes vector inversion methodology and carried out a comprehensive theoretical and numerical assessment of its key algorithms, such as polarised radiative transfer and regularisation of the ill-posed stellar surface mapping inverse problems. The forward Stokes parameter profiles computed with \inva\ were compared with the spectra produced by two other independent magnetic spectrum synthesis codes \citep{wade:2001}. This study showed a very good agreement of the local and disk-integrated Stokes profiles produced by all three codes, provided they use a consistent definition of Stokes parameters and the same input model atmosphere and atomic data.

A subsequent study by \citet[][hereafter KP02]{kochukhov:2002c} presented extensive numerical experiments designed to evaluate performance of \inva. These tests demonstrated that, given high-resolution $IQUV$ observations, the magnetic inversion code is capable of correctly reconstructing abundance and magnetic field vector surface distributions simultaneously and without any prior assumptions about the large-scale magnetic field geometry. At the same time, it was found that an accurate mapping of a globally-organised magnetic field from circular polarisation alone requires an additional multipolar constraint on the possible magnetic field structure. KP02 showed that the full Stokes vector-based simultaneous magnetic and abundance inversions are generally stable and robust. It was found that the derived solutions are unique and the magnetic mapping code cannot arrive to a spurious solution which would provide a good fit to observations. Another important outcome of the MDI tests carried out by KP02 is that magnetic inversions extract information about magnetic field primarily from the rotational modulation of Stokes profiles rather than from the Doppler shifts, unlike for the conventional scalar DI, making it possible to apply MDI to very slowly rotating stars.

The comprehensive discussion of the MDI methodology implemented in \inva\ and extensive numerical tests of this code presented by PK02 and KP02 are neglected by S12, who do not mention either of these two papers. This oversight makes much of the general criticism by S12 irrelevant because many of their concerns about magnetic inversion methodology have been explicitly addressed by PK02 and KP02.

We note that the contentions of S12 about what DI inversion can or cannot do lack a reference to the basic numerical tests demonstrating robustness of their own abundance inversion code. For instance, comparing Fig.~7 and 8 of their paper, one can see that the authors have failed to recover the true distribution of abundance spots even when they used a constant magnetic field and adopted the same mean atmospheric structure for calculations of the input spectra and for the inversion. A major discrepancy between the input and reconstructed abundance maps revealed in such a simple test indicates serious problems with the DI code used by S12. Evidently, results based on the application of this untested inversion code must be considered with caution.

\subsection{Regularisation in magnetic DI}

Typically, a grid of several hundred to several thousand surface elements is employed in DI. Because even a high-quality spectroscopic data set may only represent a few hundred independent constraints on the stellar surface structure, the conventional temperature or abundance Doppler imaging problem is mathematically ill-posed (\citealt{goncharskii:1977}; \citealt*{vogt:1987}). As a consequence, a large family of maps -- each with the same large-scale structure, but differing in smaller-scale details -- is able to reproduce the observations. In fact, the surface grid is often sufficiently fine as to be able to fit the observational noise. To solve these problems a penalty (or regularisation) function is included in the optimisation procedure, serving to limit the information content (maximum entropy regularisation) or high-frequency structure (Tikhonov regularisation) of the map. It has been demonstrated that in conventional DI, for high resolution and signal-to-noise ratio data sets, these two approaches lead to maps that are essentially equivalent \citep{strassmeier:1991}.

The situation is different for magnetic DI. \citet{piskunov:2005} showed that the underlying inverse problem based on full Stokes observations is well-posed and thus has a unique solution, and is therefore not influenced by the selection of a specific type of regularisation. In this case the role of regularisation is limited to suppressing numerical instabilities induced by a sparse wavelength and phase sampling typical of real observational data. In contrast, the MDI problem based on only Stokes $I$ and $V$ is intrinsically ill-posed, leading to different results depending on the applied regularisation algorithm. For instance, \citet{brown:1991} have failed to recover a dipolar magnetic field distribution from Stokes $IV$ data with their maximum entropy ZDI code but succeeded in reconstructing topologically more complex field geometries consisting of isolated spots. 

This counterintuitive situation when mapping simpler global field topologies from Stokes $IV$ spectra is seemingly more difficult than recovering very complex fields of late-type stars \citep[e.g.,][]{donati:1999} has been addressed by KP02. They showed that the maximum entropy constraint is inappropriate for global magnetic topologies since they do not correspond to a map with a lower information content compared to a distribution of isolated spots on a homogeneous background. KP02 achieved somewhat better results with $IV$ imaging of dipolar fields using the Tikhonov regularisation and demonstrated that a fully satisfactory reconstruction of the global field topologies is possible with multipolar regularisation, which penalises a deviation of the magnetic geometry from a low-order multipolar solution. A similar magnetic inversion technique, employing a general spherical harmonic expansion, was independently developed by \citet{donati:2001a} and successfully used for the Stokes $IV$ ZDI of global fields of Ap stars \citep{luftinger:2010a}, fully convective low-mass stars \citep{donati:2006,morin:2008}, solar analogues \citep{petit:2008a}, and late-type Ap-star descendants \citep{auriere:2011}. These developments were entirely neglected by S12, who subscribed to the obsolete view of \citet{brown:1991} that global magnetic fields can be only mapped using full Stokes data.

Discussing the $IQUV$ magnetic mapping of \cvn\ by KW10, S12 insisted, citing Figs.~6 and 8 of KW10, that these inversions are unstable because \textit{``results depend too strongly on the regularisation parameter''}. But these figures show exactly the opposite, demonstrating the impact of changing the regularisation parameter (which controls the relative contributions of the penalty function and chi-square of the fit) by a staggering factor of 10. This large increase of regularisation was required to wipe out the small-scale field structures. Nevertheless, contrary to the claim by S12, the large-scale field topology, characterised by the radial field component, was very similar for the maps obtained with the optimal and artificially large regularisation, so the Tikhonov regularisation operated as expected. KW10 have made it clear that the two values of the regularisation parameter are by no means equivalent because applying the higher regularisation value yields a far worse fit to the observational data. S12 seem to have ignored these explanations and the context of KW10 inversions with different regularisation parameters, leading to their misinterpretation of the MDI results.

\subsection{Line profile modelling and interpretation of DI maps}
\label{decop}

The general goal of Doppler imaging -- reconstructing a two-dimensional stellar surface map -- is always achieved through a modelling of the rotational modulation of spectral line profiles. However, it is important to discriminate between the tasks of obtaining a map of a certain parameter characterising the stellar surface and interpreting this map in terms of a physical quantity. In the early days of chemical spot mapping of Ap stars, DI studies reconstructed local equivalent widths instead of real elemental abundances \citep{goncharskii:1983}. The majority of current DI and ZDI studies of cool active stars map brightness instead of temperature, using a coarse temperature-independent analytical description of the local line profiles \citep[e.g.,][]{donati:2003,marsden:2011}. Modern abundance DI studies of Ap stars (e.g., \citealt{kochukhov:2004e}; \citealt*{rice:2004}; \citealt{luftinger:2010a}) are normally based on the theoretical spectra calculated for different chemical composition using the same mean model atmosphere. The resulting maps are presented in the form of horizontal distributions of chemical abundance. But it is important to remember that even in this case ``elemental abundance'' is essentially a parameter controlling the local line strength and, therefore, it is the former stellar surface property that is mapped by an inversion code. Interpretation of the line strength parameter can be viewed as a separate analysis step, largely decoupled from the mapping itself. Assumptions about the atmospheric structure of the regions characterised by different metallicity made in this latter step are not necessarily undermining reconstruction of the two-dimensional maps and have no possibility of changing the topology of the main stellar surface features as long as the resulting local profiles can match those produced with a physically correct and complete model by introducing an offset of chemical abundance from its true value.

If a magnetic field is reconstructed simultaneously with chemical maps, there is a similar decoupling between polarisation profile modelling and interpretation of the Stokes $I$ line strength. Even if the latter is in error by a significant amount, any decent MDI code would compensate for this problem by arriving to a different local abundance, which yields the local polarisation profiles closely matching the true ones for the same magnetic field strength and orientation. One the other hand, an unaccounted variation of the continuum brightness may lead to a systematic over or underestimation of the magnetic field strength in bright, respectively dark, surface regions.

This robustness of DI with respect to errors in interpretation of the local line profiles and decoupling between magnetic and abundance mapping was ignored in the discussion by S12. Throughout their paper these authors incorrectly imply that any difference between the local profiles corresponding to an approximate versus a fully realistic treatment of the line formation will automatically translate to major surface structure artefacts, invalidating magnetic mapping. This viewpoint is fundamentally flawed because it neglects to take into account the intrinsic ability of an MDI code to compensate the line intensity errors arising from the use of an approximate relation between the local line strength and chemical composition.

\subsection{An evidence for small-scale fields in Ap stars}

Traditionally, analyses of magnetic field topologies of early-type stars relied on fitting low-order multipolar field models to the phase curves of integral observables, such as the mean longitudinal field and the mean field modulus, inferred from the low- or moderate-resolution Stokes $I$ and $V$ observations \citep{landstreet:2000,bagnulo:2002}. Extension of these studies to broad-band linear polarisation revealed significant deficiencies of the multipolar modelling approach, hinting at the presence of more complex magnetic topologies on the surfaces of Ap stars (\citealt{leroy:1995}; \citealt*{leroy:1996}). A direct comparison of the multipolar field geometry predictions with the high-resolution MuSiCoS four Stokes parameter observations of Ap stars \citep{wade:2000b,bagnulo:2001} demonstrated a major failure of these topologically simple models in reproducing the shapes and amplitudes of the observed Stokes $Q$ and $U$ profiles. 

Inversions based on these pioneering full Stokes vector observations showed that small-scale field structures superimposed on the global bipolar background are  required to match the linear polarisation inside spectral line profiles (\citealt{kochukhov:2004d}; KW10). The KW10 analysis of \cvn\ provided the most clear-cut example of this situation. The MDI based on the Stokes $I$ and $V$ subset of the four Stokes parameter observations yielded a dipolar-like field geometry. On the other hand, extending the inversion to all four Stokes parameters revealed a pair of high-contrast magnetic spots superimposed on the dipolar-like background. As discussed above, this is definitely not a regularisation artefact as the difference between the linear polarisation profiles corresponding to the magnetic models with and without the small-scale features is highly significant. 

The claim by S12 that \textit{``a mere 3 out of 20 phases are responsible''} for this complexity of \cvn\ magnetic maps is false. KW10 have unambiguously stated in the text of their paper and clearly illustrated with Fig.~5 that the discrepancy between the ``simple'' and ``complex'' magnetic field geometries is large in comparison to the observational noise over at least half of the stellar rotational cycle, corresponding to approximately 10 phases. The model lacking small magnetic spots yields a substantially higher chi-square and cannot even qualitatively reproduce the double-wave shape of the $Q$ and $U$ profiles observed for the 0.3--0.6 phase interval. Below we show that for Stokes $QUV$ this difference is actually considerably more important than the impact of using individual local model atmospheres, as urged by S12.

S12 objected the very possibility of resolving small-scale surface features in relatively slowly-rotating stars like \cvn\ ($v_{\rm e} \sin i$\,=\,18.5~\kms) on the grounds that the difference in Doppler shifts between the leading and trailing edges of such spots is small in comparison to the instrumental resolution. According to them, the spatial resolution of the stellar DI images is entirely determined by the relation of $v_{\rm e} \sin i$ and instrumental profile width. Any reader familiar with stellar surface mapping will immediately recognise an obvious flow in this reasoning. S12 consider the surface resolution provided by \textit{a single observation}, while DI makes use of \textit{many time-resolved spectra} (20 in the case of KW10 study of \cvn), extracting information not only from the Doppler shifts but also from rotational modulation. The latter is more important for magnetic mapping since polarisation profiles can exhibit a huge rotational variability even if $v_{\rm e} \sin i$ is negligible. Numerical experiments by KP02 have demonstrated that this allows an application of MDI to all types of early-type magnetic stars, regardless of their projected rotational velocities. Indeed, magnetic DI maps have been obtained for many very slowly rotating ($v_{\rm e} \sin i$\,$\la$\,5~\kms) early-type \citep{donati:2006b,luftinger:2010} and late-type \citep{petit:2008a,morin:2008,fares:2010} stars. A reader may also note that with systematic application of the reasoning of S12, one is bound to conclude that stellar surface mapping is completely unfeasible with photometric data -- a ludicrous assertion given the well-known success of many photometric star spot investigations (e.g., \citealt*{korhonen:2002}; \citealt{lanza:2009,mosser:2009,luftinger:2010a}). For instance, the high photometric stability of the CoRoT satellite allowed \citet{mosser:2009} to detect photometric signatures of spots as small as a few degrees across, whereas according to S12 such a study should not have provided any surface spatial resolution at all.

Summarising, the general arguments by S12 against the detectability of the complex magnetic fields on Ap stars are demonstrably incorrect since they neglect some of the very basic DI principles and ignore a large body of contrary evidence in the literature. 

\section{Model atmospheres and local profiles}
\label{local}

\subsection{Atmospheric modelling with the LLmodels code}
\label{llmodels}

Any meaningful analysis of the effects of non-solar chemical composition on the atmospheric structure of Ap stars requires an advanced opacity sampling model atmosphere code, which allows to carry out accurate calculations for arbitrary elemental abundances. In this study we use version 8.8 of the well-established model atmosphere code \llm\ \citep{shulyak:2004}. This program treats atomic line opacity by a direct, line-by-line spectrum synthesis and is able to account for individual non-solar chemical abundance patterns \citep{khan:2007} and inhomogeneous vertical distribution of elements \citep{kochukhov:2009a,shulyak:2009}. The code also includes a provision for treating the effects of a strong magnetic field on the line opacity, including Zeeman splitting and polarised radiative transfer \citep*{kochukhov:2005a,khan:2006a}, and a contribution of the Lorentz force to the equation of hydrostatic balance \citep{shulyak:2007}. However, as demonstrated by the previous studies, for the moderate 2--4~kG magnetic field of \cvn, Zeeman splitting and polarised radiative transfer can be safely neglected in the model atmosphere calculations.

\begin{figure}
\centering
\fifps{7.0cm}{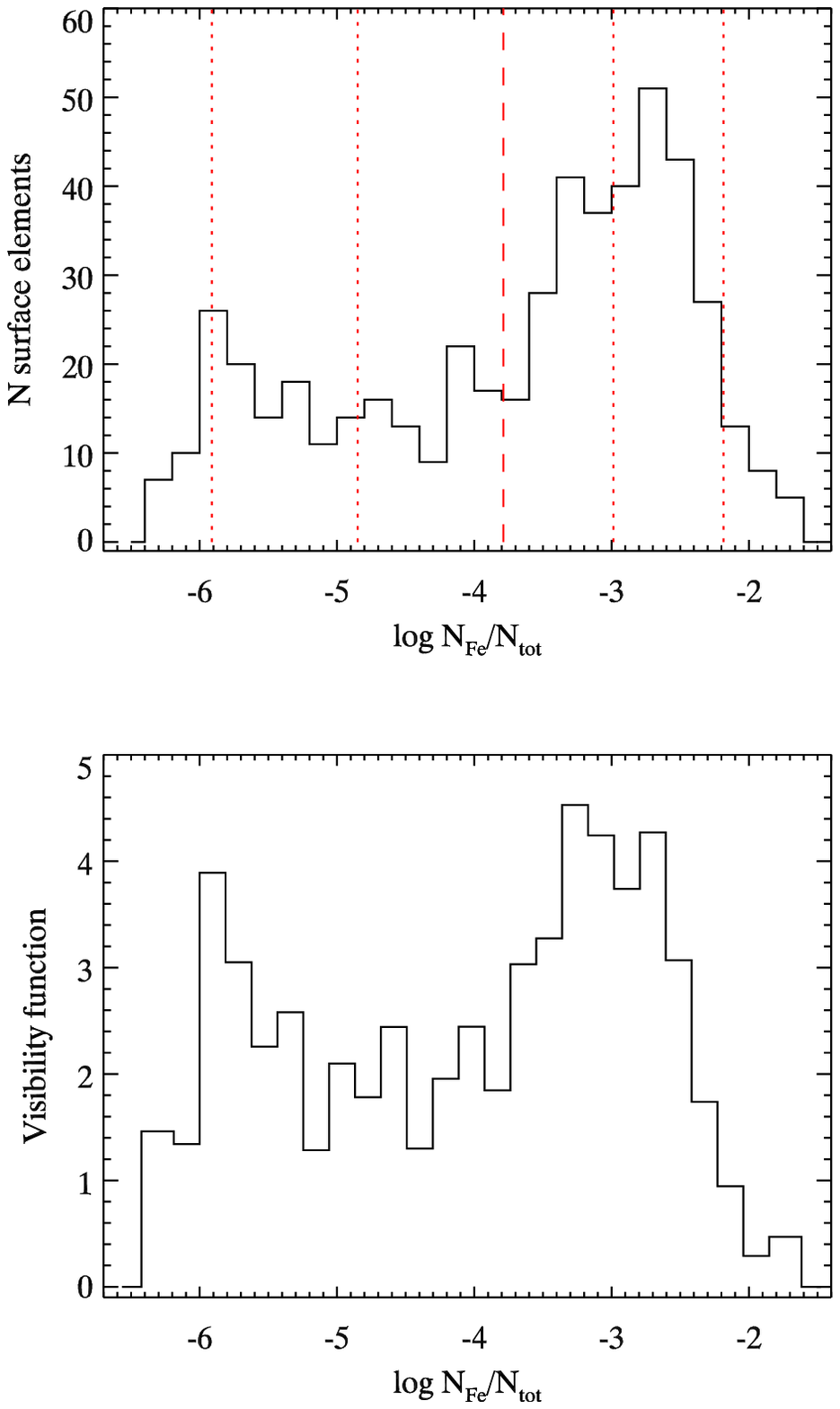}
\caption{Distribution and visibility of the Fe abundance values in the surface map derived for \cvn\ by KW10. \textit{Upper panel}: Fe abundance for the visible part of the stellar surface (latitude $\le$\,60\degr). The vertical dashed line shows the weighted mean abundance, \lgfe{-3.79}, used for the reference model atmosphere. The dashed lines indicate four additional Fe abundances for which we also calculated model atmospheres. \textit{Lower panel}: Visibility function (see Section~\ref{llmodels}) illustrating relative contributions of the surface zones with different Fe abundances to the disk-integrated spectra of \cvn.}
\label{fig:abn}
\end{figure}

Using the final Fe abundance map derived for \cvn\ by KW10, we determined a mean surface-integrated Fe abundance of \lgfe{-3.79} using a weight function according to the Eq.~(64) of PK02. The full amplitude of the Fe abundance variation over the stellar surface is 4.9~dex, but the largest values in this range are represented by a small number of surface elements and make insignificant contribution to the disk-integrated Stokes profile spectra. Figure~\ref{fig:abn} shows a distribution of the Fe abundance values for the visible part of the stellar surface, which corresponds to the latitude range from $-90\degr$ to +60\degr\ for the inclination angle $i=120\degr$ adopted for \cvn. Most of the discussion by S12 is focused on model atmospheres computed for extreme Fe overabundances, such as \lgfe{-2.0}, $-1.75$, and $-1.5$. However,  the reader can readily see from Fig.~\ref{fig:abn} that these high abundances are quite rare. In the actual Fe map published by KW10 only 2.5\% of the visible surface has $\log N_{\rm Fe}/N_{\rm tot}\ge-2$. 

\begin{figure}
\centering
\fifps{7.5cm}{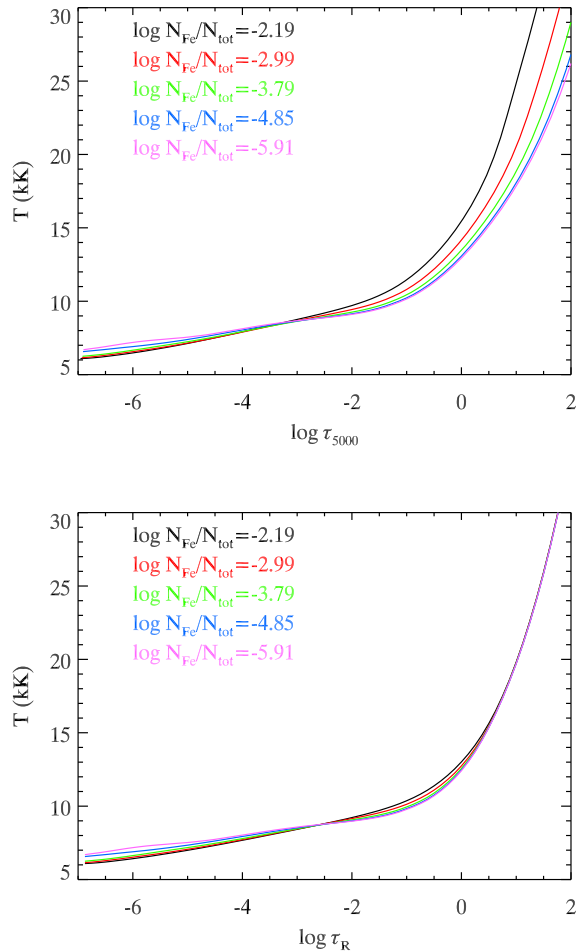}
\caption{Temperature as a function of the continuum optical depth at $\lambda=5000$~\AA\ (\textit{upper panel}) and Rosseland optical depth (\textit{lower panel}) in the model atmospheres of \cvn\ computed for different Fe abundances.}
\label{fig:ttau}
\end{figure}

In fact, the zones with a high Fe concentration are even less important for the disk-integrated Stokes spectra because they are mostly located on the stellar hemisphere directed away from the observer. The lower panel of Fig.~\ref{fig:abn} quantifies this statement by plotting a visibility function of different Fe abundances, defined as the projected areas of the corresponding surface elements summed over all 20 rotational phases of the \cvn\ observations analysed by KW10. If we normalise the area under this visibility function to 1, the relative contribution of zones with $\log N_{\rm Fe}/N_{\rm tot}\ge-2$ is only $1.1\times10^{-2}$. Thus, contrary to the claims by S12, regions of extreme Fe overabundance contribute negligibly to the MDI analysis of \cvn.

Aiming to present a balanced analysis of the effects of metallicity on the model atmosphere and spectra of \cvn, we considered the Fe abundance range from \lgfe{-2.19} to \lgfe{-5.91}, encompassing 90\% of the visible stellar surface. We have covered this range with five model atmosphere calculations, differing only by the adopted Fe abundance. For simplicity, the solar abundances from \citet*{asplund:2005} are assumed for all other chemical elements. For all model atmosphere calculations reported in this paper we used a 72-layer depth grid with an equidistant spacing in the $\log\tau_{5000}$ range between $-7$ and +2. The frequency-dependent quantities were calculated between 50~\AA\ and 100\,000~\AA\ with a constant wavelength step of 0.1~\AA. The resulting sampling at $\sim$\,$10^6$ frequency points yields more precise line opacities than the {\sc Atlas12} calculations at 120\,000 frequencies employed in the study by S12. \llm\ iterations were continued until the temperature change was less than 1~K at all depths. This condition also satisfied the criteria of the flux constancy and radiative equilibrium to 0.1--0.3\%. S12 admitted that their model calculations were plagued by convergence problems for high Fe abundances. We encountered no such problems with \llm.

Our computation of the line absorption coefficient was based on the atomic data for $\sim$\,$6.6\times10^7$ lines extracted from the VALD data base \citep{kupka:1999}. In addition to the standard lists of transitions between measured energy levels, this includes a new massive predicted line list calculated by R.~Kurucz\footnote{http://kurucz.harvard.edu/atoms.html} and discussed by \citet*{grupp:2009}. These spectral line data are essential for accurate opacity calculations in the context of Ap star atmospheric modelling, especially for large Fe overabundances explored in this study. It is not clear what predicted line list was employed by S12.

The distribution of temperature with optical depth for five \llm\ atmospheres computed for different Fe abundance is illustrated in Fig.~\ref{fig:ttau}. Looking at the plot of temperature as a function of $\tau_{5000}$, we see a prominent backwarming, which leads to a steeper temperature gradient and changes the temperature at $\log\tau_{5000}=0$ by 500--1500~K. These effects are less extreme in the higher atmospheric layers corresponding to the typical line-forming regions. 

It is important to realize, however, that $\tau_{\rm 5000}$ reflects changes in the model structure only at one wavelength. Thus, examining the plot of $T$ vs. $\tau_{\rm 5000}$, one can only say that for the models with different iron abundance the formation depth of the continuum flux at $\lambda=5000$~\AA\ corresponds to substantially different local temperature. But this picture and the resulting conclusions may change if one chooses a monochromatic optical depth which is weakly affected by the Fe abundance. 

A more natural way to compare model atmosphere structures is to plot physical quantities against an optical depth computed with some mean opacity coefficient. In particular, an optical depth based on the Rosseland opacity coefficient $\kappa_{\rm R}$ has a certain advantage: spectral regions with the smallest opacities transfer most of the radiative energy and at the same time contribute most to $\kappa_{\rm R}$. Examining the physical quantities of different models as a function of $\tau_{\rm R}$ allows one to trace changes of the model structure on a natural depth scale representing frequency-integrated properties of the radiation field.

The lower panel in Fig.~\ref{fig:ttau} illustrates the depth dependence of temperature for the same set of \llm\ atmospheres as mentioned above, but now as a function of $\tau_{\rm R}$. It is obvious that the effects of different iron abundance introduce much smaller changes compared to those implied by the $T$ vs. $\tau_{5000}$ plot. In particular, the temperature variation between different models is reduced to only a few hundred K. In the sub-photospheric layers $T(\tau_{\rm R})$ is very similar for all models simply because of the fundamental property of $\tau_{\rm R}$: when $\tau_{\rm R} \gg 1$ the temperature distribution asymptotically approaches that of a grey model, which, as is well known, depends only on $T_{\rm eff}$ and $\tau_{\rm R}$. Therefore, regions of the same $\tau_{\rm R}$ exhibit the same temperature.

Concluding, a huge variation in the model temperature distributions plotted against $\tau_{\rm 5000}$ should not be mistaken for an exhaustive and complete comparison of different models. A complete analysis should always include not only comparisons of the depth-dependent thermodynamic properties, but also different spectroscopic and photometric observables (such as energy distributions, hydrogen lines, photometric parameters, metallic line spectra) in order to quantify the importance of detected changes in the context of particular investigations.

Let us note in passing that modification of the atmospheric structure due to enhanced metal abundances is closely connected to a well-known effect of the redistribution of stellar flux from UV to longer wavelength regions \citep{leckrone:1973}, making the Fe spot region darker in the UV and brighter in the optical wavelengths than the surrounding atmosphere \citep[e.g., see Fig.~9 of][]{luftinger:2010a}. Contrary to the impression conveyed by S12, this prominent influence of chemical composition on the structure of stellar surface layers is well-known to the Ap star community and has been addressed in numerous applications of the \llm\ code to individual stars (\citealt{kochukhov:2009a}; \citealt*{shulyak:2008}; \citealt{shulyak:2009,shulyak:2010a,luftinger:2010a}) -- a significant and fertile research direction, seemingly unnoticed by S12.

\subsection{Local line profiles and continuum intensities}
\label{locprf}

To investigate repercussions of metallicity-induced changes of atmospheric structure on the local line shapes and continuum intensity, we computed the Stokes $IQUV$ profiles of the \fe\ 5018.44~\AA\ line using polarised spectrum synthesis code {\sc Synmast} \citep*{kochukhov:2010a}. This spectral line choice is determined by previous magnetic DI studies (K04, KW10), where this feature, along with \fe\ 4923.93~\AA, was one of the very few individual metal lines showing detectable linear polarisation signature in the MuSiCoS four Stokes parameter spectra of Ap stars \citep{wade:2000b}. As emphasised by KW10, \fe\ 4923 and 5018~\AA\ generally represent a poor choice for Doppler imaging of chemical spots because these lines are saturated in a typical spectrum of a Fe-rich Ap star and thus are weakly sensitive to the horizontal abundance variations. On the other hand, strong lines are noticeably affected by vertical abundance stratification (\citealt*{ryabchikova:2005a}; \citealt{kochukhov:2006b}), which is usually ignored in Doppler mapping. Non-magnetic abundance DI analyses \citep{kochukhov:2004e,luftinger:2010a} as well as MDI studies based on the Stokes $I$ and $V$ spectra \citep{kochukhov:2002b,luftinger:2010} normally avoid such strong lines, dealing with weak or intermediate strength features. To assess how the latter lines behave in the Fe-rich \llm\ atmospheres, we have computed a second set of Stokes profiles for a fictitious weaker iron line with the same parameters as \fe\ 5018.44~\AA, but with $\log gf$ reduced by 1~dex.

\begin{figure*}
\centering
\fifps{17cm}{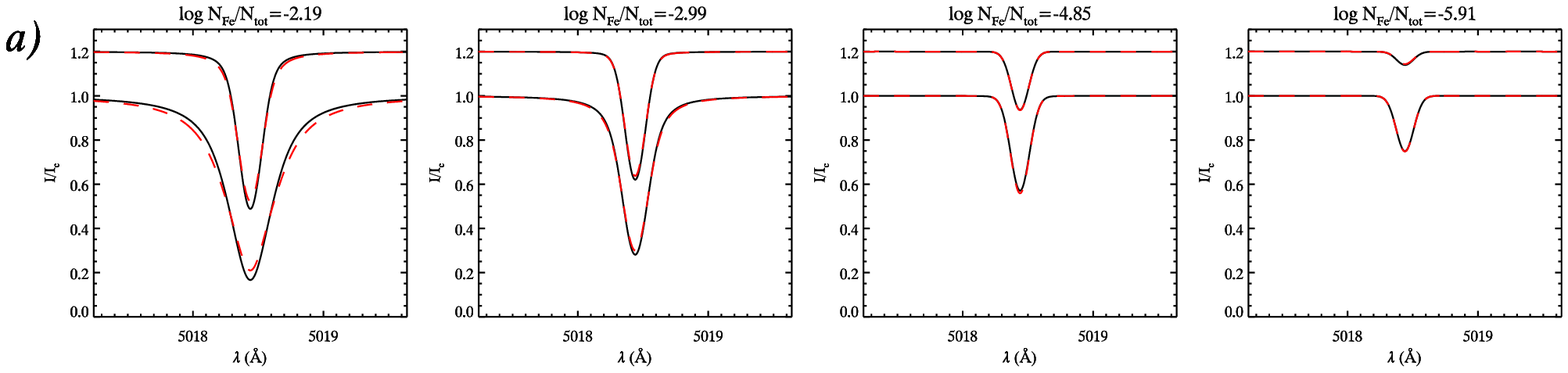}
\fifps{17cm}{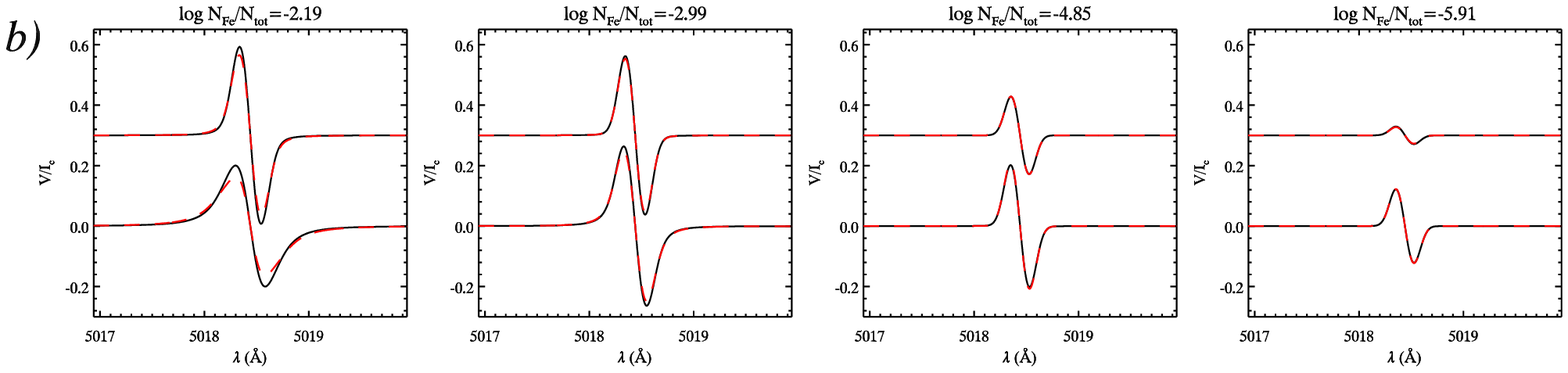}
\fifps{17cm}{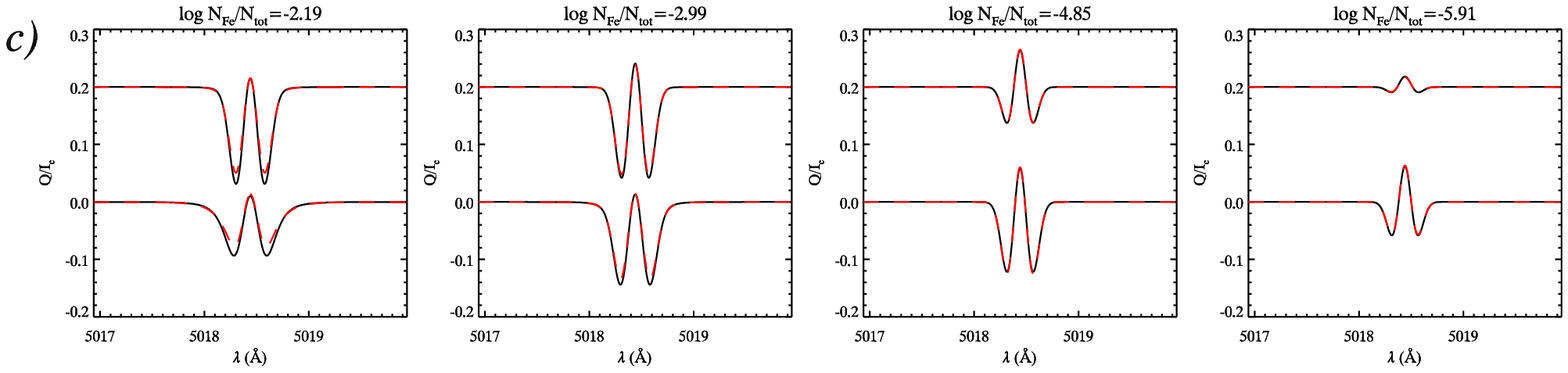}
\caption{Comparison of the normalised disk-centre profiles of the \fe\ 5018.44~\AA\ line (lower profiles in each panel) computed self-consistently for the model atmospheres with different iron abundances (\textit{solid lines}) with the  profiles calculated for the same abundances and a single mean atmosphere (\textit{dashed lines}). The same comparison is presented for a fictitious weaker \fe\ line (upper profiles in each panel). The corresponding spectra are shifted vertically by 0.2 of the continuum. The three rows of plots show calculations for {\bf a)} Stokes $I$ in the absence of magnetic field, {\bf b)} Stokes $V$ for the 3.5~kG line of sight magnetic field, {\bf c)} Stokes $Q$ for the 4.0~kG transverse magnetic field.}
\label{fig:prfIVQ}
\end{figure*}

Starting from the non-magnetic case, we show in Fig.~\ref{fig:prfIVQ}a normalised disk-centre Stokes $I$ profiles of \fe\ 5018.44~\AA\ and the weaker line for all five model atmospheres discussed above. One set of line profiles was computed self-consistently with the Fe abundance adopted for the model atmosphere calculation, while another set represents the usual approach based on a mean atmospheric model with \lgfe{-3.79} and different Fe abundances in the spectrum synthesis. These calculations were repeated for a 3.5~kG longitudinal field as well as for a 4.0~kG transverse field in order to assess any impact of the atmospheric structure on the Stokes $V$ and $Q$ profiles, respectively. The $V$ and $Q$ spectra are illustrated in Fig.~\ref{fig:prfIVQ}b and c. All calculations included convolution with a Gaussian instrumental profiles appropriate for the four Stokes parameter MuSiCoS observations employed by K04 and KW10.

\begin{figure}
\centering
\fifps{7.5cm}{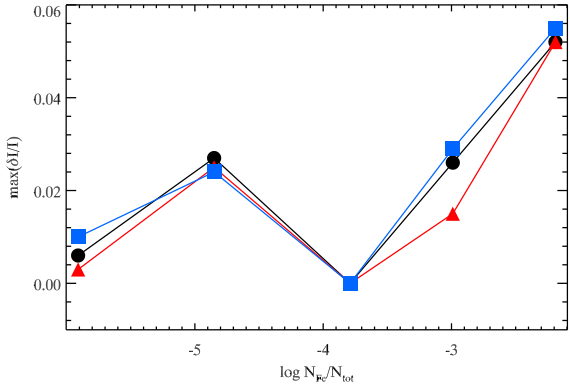}
\caption{Maximum relative profile difference of the normalised disk-centre Stokes $I$ spectra of the \fe\ 5018.44~\AA\ line calculated self-consistently with a model atmosphere structure and using a single mean model atmosphere. Symbols show calculations for zero magnetic field (\textit{circles}), 3.5~kG line of sight field (\textit{triangles}), and 4.0~kG transverse field (\textit{squares}).}
\label{fig:difI}
\end{figure}

\begin{figure}
\centering
\fifps{7.5cm}{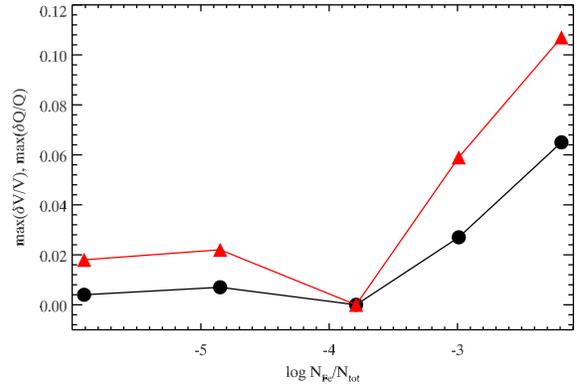}
\caption{
Same as Fig.~\ref{fig:difI} for the maximum relative profile difference of the normalised disk-centre Stokes $V$ in the 3.5~kG line of sight field (\textit{circles}) and Stokes $Q$ in the 4.0~kG transverse field (\textit{triangles}).}
\label{fig:difVQ}
\end{figure}

\begin{figure}
\centering
\fifps{7.5cm}{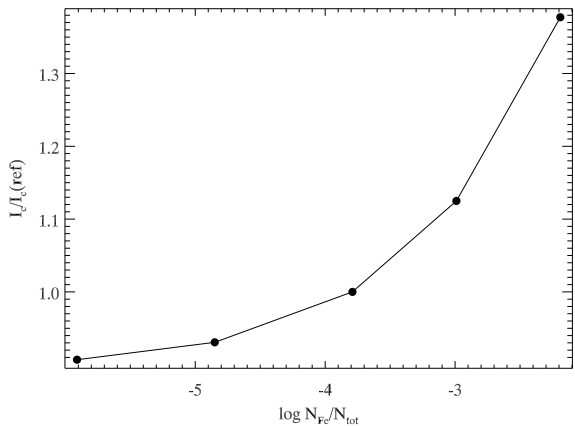}
\caption{Relative continuum intensity at $\lambda$\,=\,5018.44~\AA\ as a function of the Fe abundance adopted for the model atmosphere calculation.}
\label{fig:ic}
\end{figure}

Figures~\ref{fig:prfIVQ}a--c leave no doubt that the difference between the normalised local line profiles computed with different model atmospheres and with a mean atmosphere and different Fe abundances is extremely small. One cannot visually distinguish the two sets of calculations except for the strongest line, forming in the most Fe-rich atmosphere. Even in that case, the discrepancy is only 4\% of the continuum for the line which is 83\% deep. In general, the maximum relative difference does not exceed several \% of the Stokes $I$ profile amplitude for \fe\ 5018.44~\AA\ (Fig.~\ref{fig:difI}) and is entirely negligible for a weaker line. 

How does this difference compare with the sensitivity of Stokes $I$ to abundance variation? Considering the \fe\ 5018.44~\AA\ line for the model with \lgfe{-2.19}, we found that the Fe abundance needs to be changed by just $\approx$\,0.10~dex, or 2\% of the logarithmic Fe abundance range for \cvn, to obtain the same maximum profile difference as seen in Fig.~\ref{fig:difI}. 

The metallicity-induced modification of the atmospheric structure leads to a tiny variation of the normalised Stokes $V$ and $Q$ profiles, illustrated in Fig.~\ref{fig:prfIVQ}b and c, respectively. In the worst case scenario, these changes amount to about 0.05--0.10 of the full polarisation profile amplitude (see Fig.~\ref{fig:difVQ}). Generally, the polarisation profiles become slightly broader, while their low-order moments remain unchanged.

The full Stokes profile calculations presented in this section fully support the conclusion by KW10 that \textit{``the local line profiles are sensitive to model structure effects to a much smaller degree than to changes of abundance or magnetic field''}. At the same time, enhanced metallicity leaves a major imprint on the continuum intensity, which is increased in the optical due to the flux redistribution from the short-wavelength regions. Figure~\ref{fig:ic} demonstrates that the Fe-rich zones are expected to be up to 40\% brighter in the continuum at $\lambda$ 5018.44~\AA\ with respect to the reference model, while the regions with a relative Fe underabundance turn out to be up to 10\% fainter. The consequences of these continuum intensity changes for the disk-integrated Stokes $IQUV$ profiles and for the reconstruction of the surface abundance and magnetic field distributions will be addressed in the next section.

The general conclusion regarding the insignificance of the differential line-blanketing for the local spectral line shapes emerging from our calculations seems to be drastically different from the claims made by S12 in Section~4 of their paper. This discrepancy comes from the fact that S12 have neglected to consider the entire Fe abundance range and different line strengths, limiting their discussion to the profile of a single, very strong line computed for extreme Fe overabundances which occupy a tiny fraction of the visible surface of \cvn. In addition, their local profile analysis lacked a comparison of the intensity differences with the actual profile amplitude, as in our Fig.~\ref{fig:prfIVQ}, and did not at all address polarisation spectra. As revealed by our more comprehensive calculations, the atmospheric structure variations result in a systematic change of the continuum fluxes but has a minuscule influence on the normalised intensity and polarisation lines shapes, except for a small region of the parameter space (the strongest line in the most Fe-rich atmosphere). As discussed above, we do not expect this region to contribute significantly to the disk-integrated line profiles. This is directly confirmed by the spectrum synthesis calculations presented in the next section.

\section{Magnetic DI with self-consistent model atmosphere structure}
\label{invers13}

\subsection{Magnetic inversion code Invers13}

We have carried out a series of forward line profile calculations to assess the impact of the anomalous local model atmosphere structure on the disk-integrated Stokes spectra of \cvn. For these experiments we used our new magnetic DI code \invb. This efficient and versatile code represents a further development of our magnetic inversion tool \inva, which has been described in detail by PK02, evaluated by \citet{kochukhov:2002c}, and was subsequently applied in several studies of a non-uniform distribution of magnetic fields and chemical abundances on the surfaces of Ap stars \citep[K04, KW10,][]{luftinger:2010,kochukhov:2011a}. 

\invb\ and \inva\ adopt similar regularisation and non-linear optimisation strategies, based on the Tikhonov regularisation and modified Levenberg-Marquardt method, respectively. However, instead of using a single model atmosphere structure as was done in all previous DI studies of Ap stars, \invb\ is capable of mapping the three components of the magnetic field vector and treating simultaneously one additional scalar surface map (temperature, abundance of a given element, magnetic field strength) fully self-consistently with the local model atmosphere structure. The basic methodology of applying \invb\ to Ap stars is thus very similar to a self-consistent magnetic and temperature mapping of cool active stars, except that the parameter distinguishing different local model atmospheres is a chemical abundance instead of \teff.

At the start of calculations, a grid of model atmospheres (either \mbox{\sc Atlas}, {\sc Marcs} or {\sc LLmodels}) is processed by \invb. The equations of chemical equilibrium (cool atmospheres) or ionisation balance (hot atmospheres) are solved with the help of a robust molecular equilibrium solver \citep*{valenti:1998} implemented in {\sc SME} \citep{valenti:1996} and in the {\sc Synth3}/{\sc Synthmag}/{\sc Synmast} family of codes \citep{kochukhov:2007d,kochukhov:2010a}. Then, \invb\ calculates the depth-dependent line and continuum opacity coefficients for all model atmospheres and all molecular and atomic spectral lines included in the input list. The atomic line parameters are taken from VALD \citep{kupka:1999}, whereas the data on molecular transitions are taken from the {\sc MARCS} opacity database \citep{gustafsson:2008}.

The core module of \invb\ is the quadratic DELO algorithm for highly accurate solution of the LTE polarised radiative transfer equation in an arbitrary stellar atmosphere permeated by a depth-independent magnetic field with a given (local) strength and orientation. PK02 demonstrated that the quadratic DELO formal solver is superior in terms of the speed and accuracy to the Zeeman Feautrier method used by S12. For each element in a stellar surface grid optimally sampling different latitudes (see Fig.\,5 in PK02), our code evaluates the line absorption matrix by adding opacities of all contributing Zeeman components and computes the local Stokes $IQUV$ profiles and continuum intensity $I_{\rm c}$ by performing a quadratic interpolation within three sets of model spectra corresponding to the model atmospheres bracketing the value of a scalar parameter. This also yields precise derivatives of the Stokes $IQUV$ spectra and associated intensity $I_{\rm c}$ with the respect to a scalar parameter distinguishing different models in the input model atmosphere grid. One-sided derivatives with respect to the radial, meridional, and azimuthal magnetic field vector components are computed numerically. The full calculation of the local Stokes spectra and their derivatives thus consists of 12 separate polarised line profile calculations for every surface element at every relevant rotational phase.

The local profiles are convolved with a Gaussian function to model instrumental and radial-tangential macroturbulent broadening. Profiles are Doppler-shifted and summed with appropriate weights for each rotational phase. A similar phase-dependent disk-integration is carried out for the continuum intensity. The resulting Stokes $IQUV$ spectra are normalised by the respective continuum fluxes. The matrix of partial derivatives of every observed spectral pixel with respect to every surface parameter is established taking into account both the magnetic and model structure sensitivity of the line profiles and the model atmosphere sensitivity of the continuum fluxes. Using these derivatives, the code establishes the curvature matrix required by the Levenberg-Marquardt algorithm, adds regularisation terms and iteratively adjusts surface maps to match the observations.

\invb\ takes advantage of the standard MPI libraries available at any modern multi-CPU computer to perform highly efficient parallel radiative transfer calculations with a novel master-slave technique discussed by PK02. We usually run the code with one master process collecting results of the radiative transfer calculations by 16--32 slave processes. The code was successfully tested, showing a linear scaling, with up to several hundred CPUs in simultaneous inversions of over a thousand atomic and molecular lines spread over ten wavelength intervals. In practice, the number of spectral lines and wavelength intervals modelled simultaneously is limited only by the available RAM. For the calculations described here, modelling of one or two \fe\ spectral lines at 20 rotational phases required less than one minute of computing time for the forward spectrum synthesis and about three hours for a complete inversion on a modest, 10-year-old 16-CPU computer. This experience shows that currently available computing resources and standard scientific programming techniques based on well-established {\sc Fortran} codes enable a fully realistic treatment of the lateral variation of stellar atmosphere, without a recourse to the next generation  computers or exotic programming languages advocated by S12.

\invb\ was rigourously tested with numerical experiments for cool active stars using simulated circular polarisation and full Stokes vector data \citep{kochukhov:2009c}. The code was also applied to interpret circular polarisation observations of the active RS~CVn star II\,Peg \citep{kochukhov:2009d}. A comprehensive MDI study of this target is currently underway.

\begin{figure}
\centering
\figps{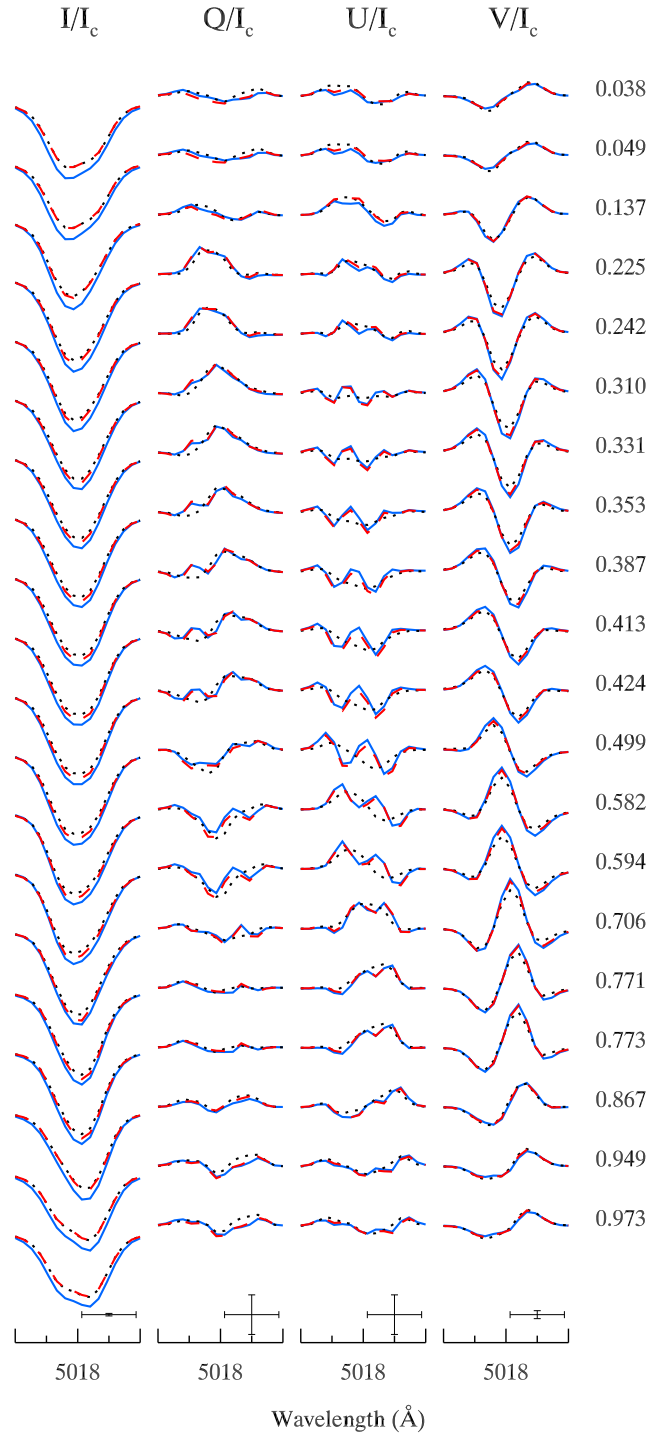}
\caption{Theoretical four Stokes parameter profiles of the \fe\ 5018.44~\AA\ line corresponding to the final magnetic and abundance maps derived for \cvn\ by KW10. This plot compares the self-consistent profiles (\textit{solid line}) with the standard calculations using a mean model atmosphere (\textit{dashed line}). For comparison, we also show profiles for the smoothed magnetic field geometry model discussed by KW10 (\textit{dotted line}). Spectra for consecutive rotational phases are shifted vertically. Rotational phases are indicated on the right. The bars at the lower right of each column show the horizontal and vertical scale (0.5~\AA\ and 1\% of the Stokes $I$ continuum intensity respectively).}
\label{fig:cvn_prf}
\end{figure}

\begin{figure}
\centering
\figps{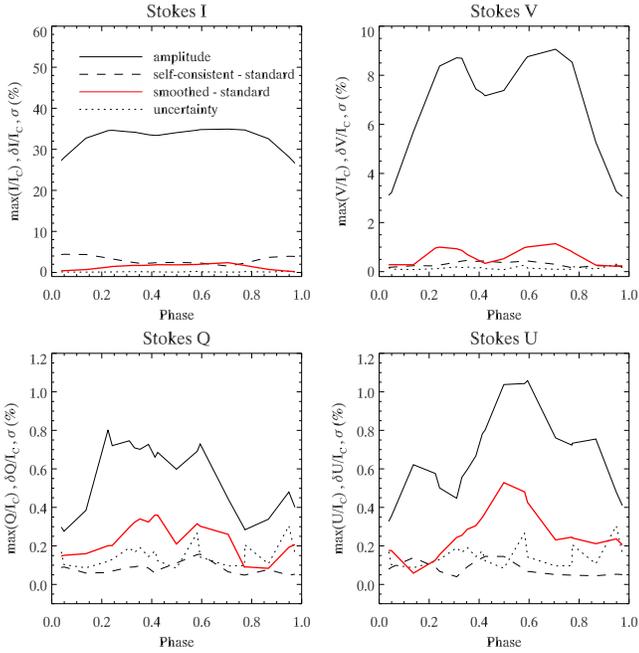}
\caption{Phase dependence of the amplitude, observational uncertainty, and the difference between theoretical Stokes parameter profiles illustrated in Fig.~\ref{fig:cvn_prf}. This plot compares the amplitude of Stokes profiles (\textit{black solid line}) with the discrepancy between the standard and self-consistent calculation (\textit{dashed line}), and observational uncertainty (\textit{dotted line}). For comparison, we also show the difference between the standard profiles and the spectra corresponding to the smoothed model of the magnetic field topology (\textit{red solid line}).}
\label{fig:cvn_dif}
\end{figure}

\subsection{Disk-integrated $IQUV$ profiles of the \fe\ 5018~\AA\ line}
\label{global}

We have used the forward polarised spectrum synthesis calculations with \inva\ and \invb\ to directly answer the question of how important is the local atmospheric structure for the KW10 analysis of the Fe abundance distribution and magnetic field geometry of \cvn. We adopted the magnetic and abundance maps derived in that study and employed a single {\sc Atlas9} model atmosphere with parameters \teff\,=\,11600~K, \lgg\,=\,3.9 and $[M/H]=+1.0$ for the full Stokes parameter synthesis of the \fe\ 5018.44~\AA\ line with \inva. These calculations are identical to the final Stokes profile fits presented by KW10. In particular, we consider the same set of rotational phases and take into account the spectral resolution of the MuSiCoS Stokes vector data available for \cvn.

Another set of line profiles was produced with the \invb\ code using the grid of {\sc LLmodels} atmospheres discussed in Section~\ref{llmodels}. To avoid extrapolation in Fe abundance, we augmented this grid with two extreme models computed for \lgfe{-1.60} and $-6.40$. We checked that using a denser grid of 13 model atmospheres makes no difference. The resulting normalised \inva\ and \invb\ Stokes parameter profiles are compared in Fig.~\ref{fig:cvn_prf}. In addition, we plot a set of synthetic \inva\ Stokes $IQUV$ parameters corresponding to the smoothed model of the magnetic field geometry of \cvn\ discussed by KW10. A detailed assessment of the difference between the ``standard'', ``self-consistent'' and ``smoothed'' model profiles with respect to the observational uncertainties is presented in Fig.~\ref{fig:cvn_dif}. We note that for polarisation this comparison represents a conservative estimate of possible errors because we do not allow the code to adjust the Stokes $I$ profile intensity as it would do in a real magnetic inversion (see discussion in Section~\ref{decop}), thereby improving the fit to $QUV$.

What can we learn from this comparison? Clearly, the effect of self-consistent calculations is noticeable for Stokes $I$. The intensity profile changes by 2--4\% of the continuum, which is not negligible compared to the central line intensity of 27--35\%. In agreement with the prediction made in Section~\ref{locprf}, the self-consistent intensity profiles are systematically deeper due to an increased continuum brightness of the Fe-rich surface regions. However, this effect can be compensated by a mere $\sim$\,0.1~dex reduction of the average Fe abundance, so it is not immediately obvious that accounting for the local atmospheric structure should lead to any major change of the abundance inversion results and remove extremely high abundances from the Fe map.

\begin{figure*}
\centering
\fifps{16cm}{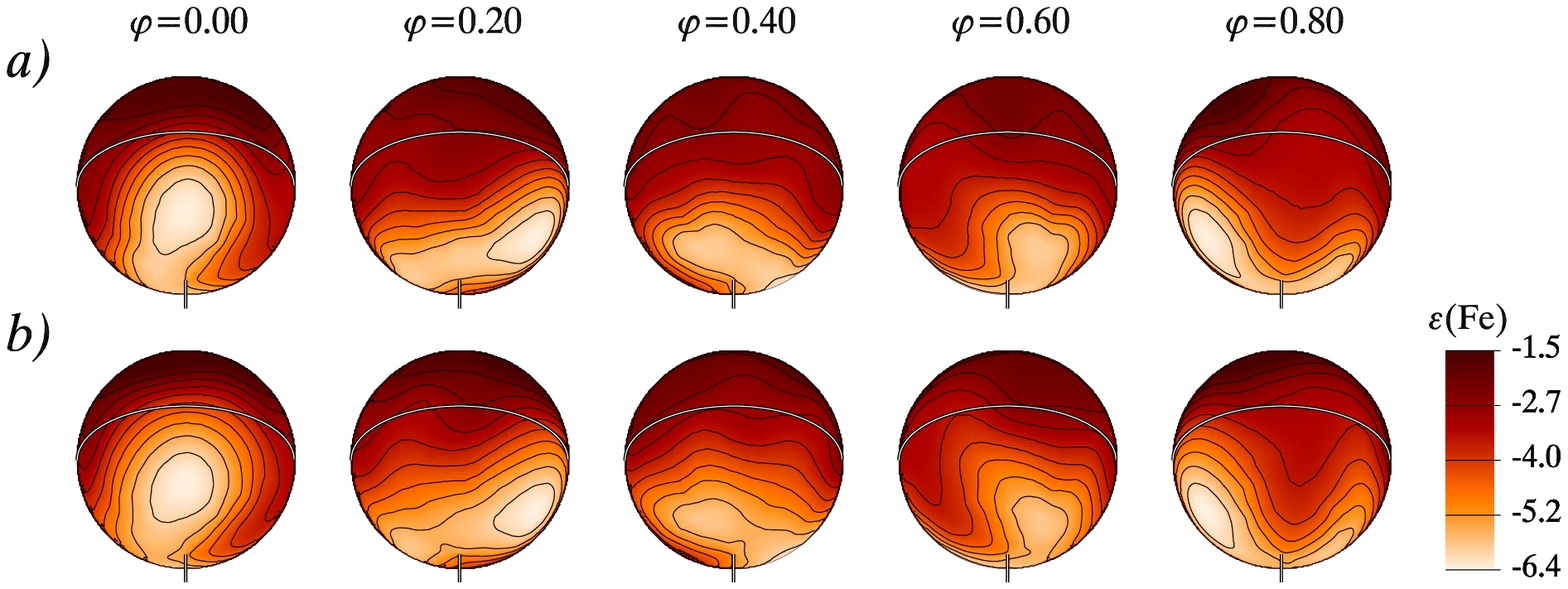}
\fifps{16cm}{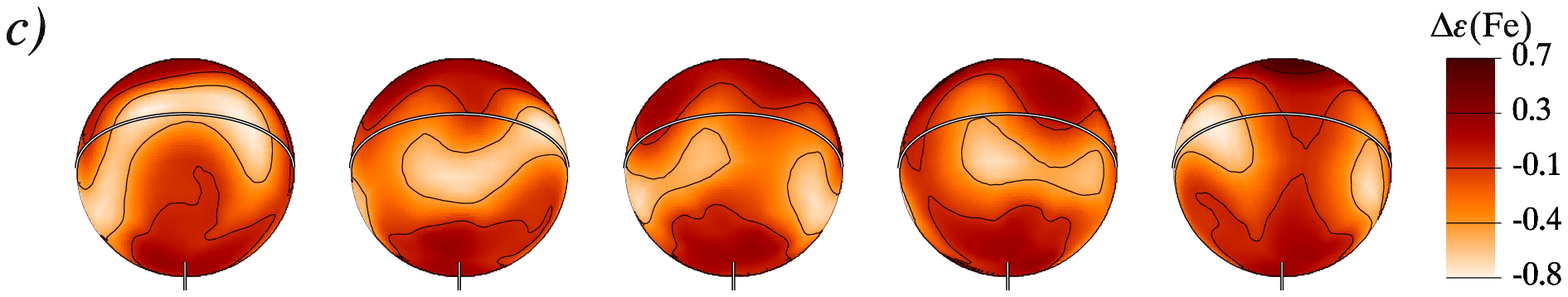}
\caption{Surface distribution of Fe abundance derived for \cvn\ from the Stokes $IQUV$ profiles of the \fe\ 4923.93 and 5018.44~\AA\ lines. {\bf a)} Inversion with \inva\ using a single model atmosphere (KW10), {\bf b)} inversion with \invb\ using a self-consistent treatment of the Fe abundance and model atmosphere structure (this study), {\bf c)} the difference map. The spherical maps are shown for five equidistant rotational phases as indicated on the top of the figure. The aspect corresponds to the inclination angle $i=120\degr$ and vertically oriented rotational axis. The contour lines are plotted with a step of 0.5~dex. The stellar rotational equator is shown with a thick line, while the vertical bar indicates rotational axis.}
\label{fig:map_abn}
\end{figure*}

\begin{figure*}
\centering
\fifps{16cm}{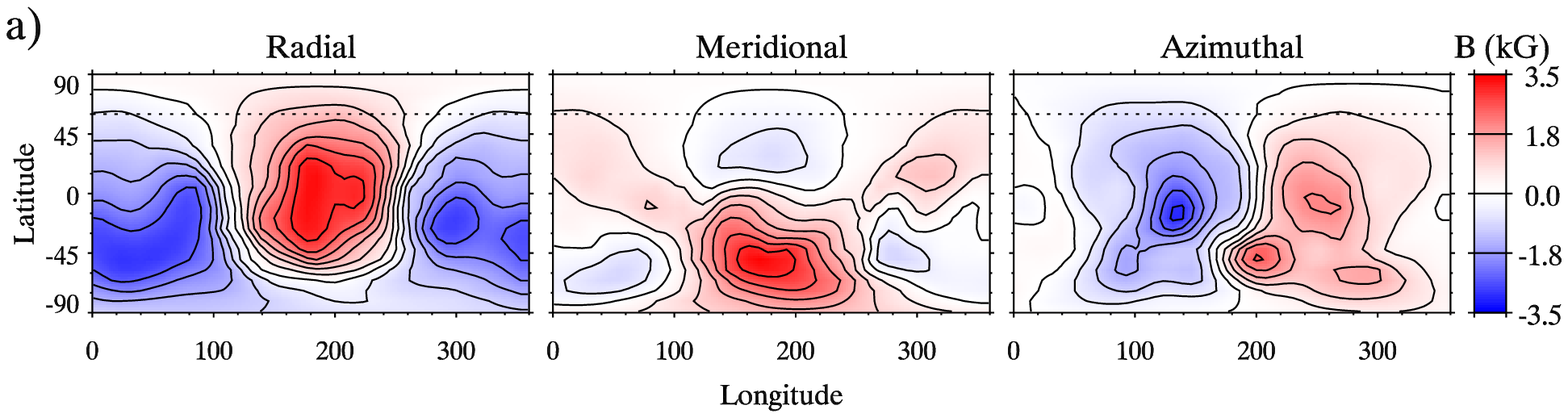}
\fifps{16cm}{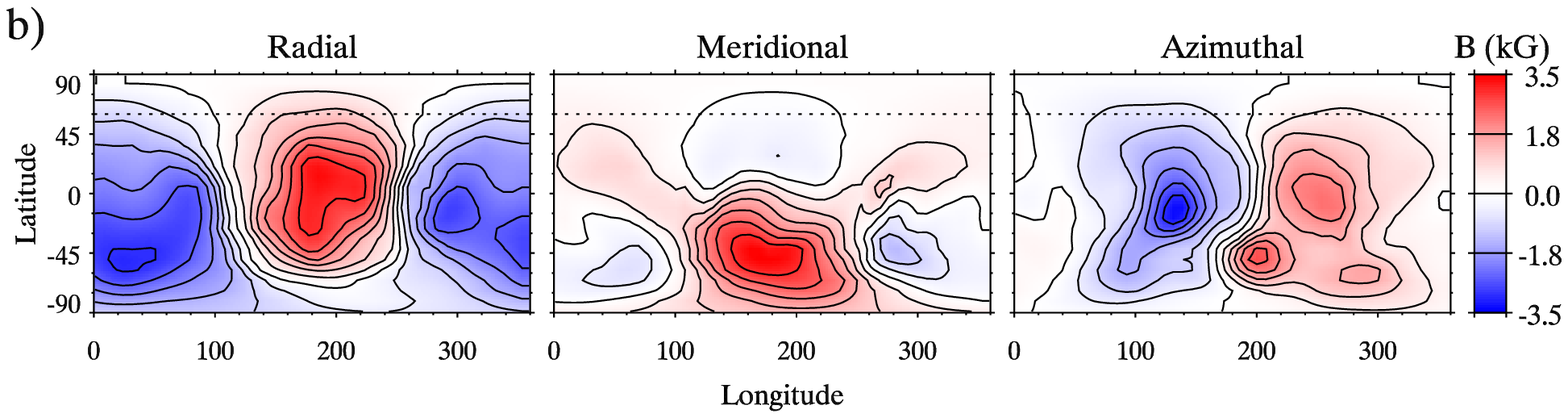}
\fifps{16cm}{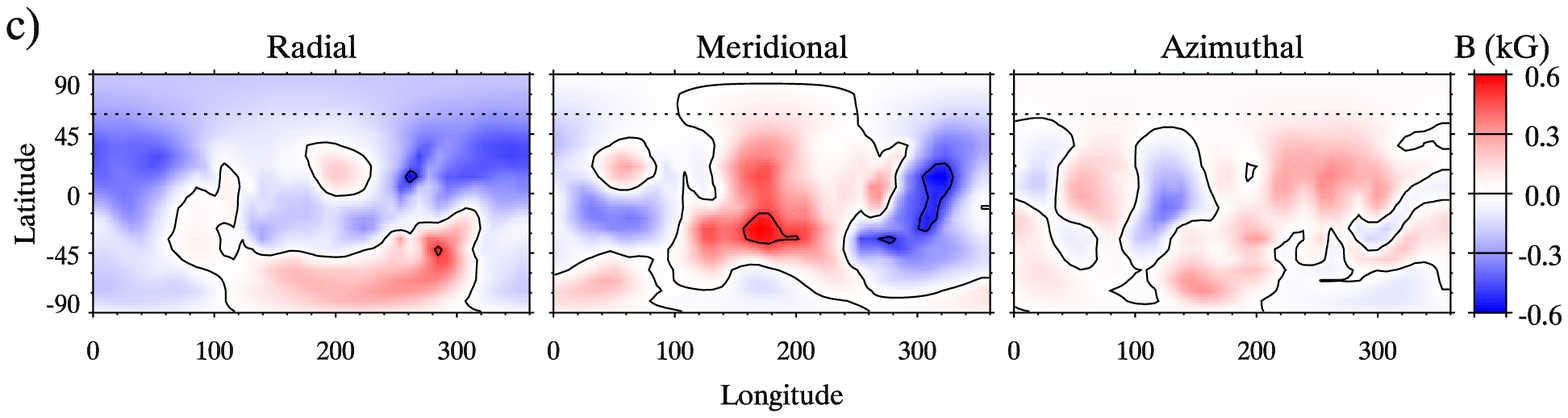}
\caption{Rectangular maps of the radial, meridional, and azimuthal vector components of the magnetic field structure derived for \cvn\ from the Stokes $IQUV$ profiles of the \fe\ 4923.93 and 5018.44~\AA\ lines. {\bf a)} Magnetic inversion with \inva\ using a single model atmosphere (KW10), {\bf b)} magnetic inversion with \invb\ using a self-consistent treatment of the Fe abundance and model atmosphere structure (this study), {\bf c)} the difference map. The dotted line shows the highest visible latitude for the inclination angle $i=120\degr$. The contours of equal field strength are plotted with a 0.5~kG step.}
\label{fig:map_fld}
\end{figure*}

At the same time, it is striking to see how little effect the use of self-consistent modelling of the Stokes profiles has on the linear and circular polarisation spectra. The discrepancy in Stokes $V$ is only 0.2--0.4\%, which exceeds the observational uncertainties but is negligible compared to the maximum circular polarisation signal of 9\%. The impact on the linear polarisation spectra is even smaller. It is evident from Fig.~\ref{fig:cvn_dif} that accounting for the local atmospheric structure changes Stokes $Q$ and $U$ by about 0.1\%, which is below the noise level of observations for all but two rotational phases.

It is enlightening to compare the difference between the standard \inva\ calculation and self-consistent \invb\ model with the discrepancy between the former spectra and those corresponding to the smoothed version of the magnetic field topology of \cvn\ (thick solid line in Fig.~\ref{fig:cvn_dif}). We remind the reader that this discrepancy, constituting the main evidence for the small-scale magnetic fields in \cvn, was dismissed by S12 as insignificant. Figures~\ref{fig:cvn_prf} and \ref{fig:cvn_dif} leave no doubt that the conclusions of S12 are erroneous. A signature of the small-scale fields yields a systematic modulation of the linear polarisation spectra, which is above the observational noise for 15 out of 20 rotational phases in Stokes $Q$ and for 18 phases in $U$. We see that the actual situation is just the opposite of that claimed by S12 in Sections~5 and 7 of their work: the influence of structured fields is highly significant for the Stokes $VQU$ profiles, while the assumption of a mean model atmosphere is largely unimportant for $V$ and completely negligible for $QU$ even in this pessimistic assessment (no abundance adjustment, a strong line). 

The results of the forward full Stokes profile calculations unequivocally demonstrate that the speculations by S12 about the significance of the local model atmosphere structure for the modelling of polarisation in \cvn\ are incorrect. At this point the reader is certainly entitled to ask what led S12 to these erroneous conclusions. The Fe distribution considered by S12 in Section~5 of their paper appears to be far more extreme ($\log (N_{\rm Fe}/N_{\rm tot}) \ge -2$ over 18\% of the visible stellar surface right at the disk centre) compared to the one found for \cvn\ by KW10 (only 2.5\% of the visible surface close to the limb). The unspecified ``eccentric dipole'' with an unusually high contrast (field strength variation by a factor of 3.7 instead of a factor of 2 expected for a centred dipole) seems to have been hand-picked to maximise profile differences. Curiously, describing the Stokes parameter calculations for this large Fe spot and unusual magnetic geometry, S12 do not mention taking into account the instrumental profile of the spectrograph, although they stress a moderate resolution of the MuSiCoS four Stokes parameter data elsewhere in the paper. It is also very surprising that the $Q$ and $U$ profiles, for which the effect of self-consistent modelling is the least important but which conversely have played a key role in the discovery of small-scale magnetic features on the surface of \cvn, are completely neglected in that part of the S12 paper. 

Calculations presented in Section 7 of the S12 paper are similarly inadequate. In that part of their study the authors claimed to have assessed the feasibility of detecting a representative signal of a high-contrast magnetic spot. However, the scenario considered is not applicable to the case of \cvn\ because the adopted spot diameter is underestimated by a factor of two and the imposed correlation of the small-scale abundance and magnetic field enhancements is contrary to what was found for that star. Furthermore, S12 produced a small-scale magnetic feature by simply scaling all three vector components in an arbitrarily chosen location on the star instead of introducing horizontal field structures similar to those found for \cvn\ by KW10. These pitfalls notwithstanding, the authors obtained differential polarisation signatures of up to $\sim$0.2\% (for a spectral resolution of $R=35000$). These signatures are about twice lower than in the real star (see $Q$ and $U$ panels in Fig.~\ref{fig:cvn_dif}), yet could still be detected with the help of multiple spectral lines and multiple Stokes profile observations at $S/N$\,$\sim$\,$10^3$. But this profoundly multi-line, multi-phase character of modern MDI was not considered by S12, who limited their analysis to a single spectral line at one rotational phase.

In summary, our forward Stokes profile calculations demonstrate a moderate metallicity effect of the local model atmosphere on the Stokes $I$ profile of one of the strongest Fe lines in the optical spectrum of \cvn, but very small ($V$) or negligible ($QU$) influence on other Stokes parameters. Local Stokes parameter calculations (Section~\ref{locprf}) suggest that the corresponding effect on weaker lines will be negligible for all four Stokes parameters. 

\subsection{Self-consistent inversion of the Fe abundance and magnetic field structure}
\label{inversion}

Taking advantage of the capabilities of \invb, we have carried out a complete reconstruction of the Fe spot distribution and magnetic field geometry on the surface of \cvn\ using the full Stokes vector observations by KW10. Incidentally, this is the first ever attempt to directly incorporate variation of the stellar atmospheric structure in the Doppler imaging of an Ap star. We modelled simultaneously the Stokes $IQUV$ profiles of the \fe\ 4923.93 and 5018.44~\AA\ lines, keeping the same strength of the Tikhonov regularisation and adopting the same stellar parameters as employed by KW10 for their final full Stokes vector MDI inversion. The iron abundance was treated self-consistently with the model atmosphere structure using the {\sc LLmodels} grid described above. A comparison of the resulting DI map with an outcome of the previous \inva\ modelling based on a single mean model atmosphere provides the most direct and conclusive assessment of the possible adverse influence of the usual simplified treatment of the Ap-star atmospheric structure. We note that the claims of S12 are not supported by a similar assessment because their inversion code \citep{stift:1996a} is not capable of magnetic mapping and, at the moment, cannot account for the lateral variation of atmospheric structure.

A comparison of the Fe abundance maps reconstructed by \inva\ and \invb\ from the same observational material is presented in Fig.~\ref{fig:map_abn}. Clearly, the chemical spot distributions recovered with the two methods are very similar and there is no sign whatsoever of a major re-arrangement of spots or significant reduction of the abundance contrast anticipated by S12. The main Fe features are located in the same regions of the stellar surface and show consistent amplitudes. We have to turn our attention to a difference map (Fig.~\ref{fig:map_abn}c) for a closer inspection of the discrepancy between the two inversions. Generally, \invb\ yields a somewhat lower Fe abundance with respect to \inva. The mean difference is $-0.12$~dex, while the standard deviation is 0.33~dex. As evident from the difference map, the discrepancy reaches $\sim$\,0.5~dex over a significant fraction of the visible stellar surface. The extreme outliers, occupying a few surface pixels, are showing abundance differences at the level of $\pm0.7$~dex. Although this change of the chemical map is not negligible, it still corresponds to only 10--15\% of the full logarithmic range of the Fe abundance. At the same time, the mean abundance changes by just 2\% of that range.

The application of \invb\ to \cvn\ clearly demonstrates that a less detailed treatment of the atmospheric structure by \inva\ does not introduce spurious abundance structures and recovers basically the same Fe horizontal abundance contrast as a more sophisticated modelling. We are far from insisting that these Fe spot mapping results are completely free from systematic errors. But it is obvious that a mean atmosphere assumption is neither a dominant nor a significant source of uncertainty for the published MDI studies of \cvn\ and other Ap stars. In particular, this assumption is definitely not the reason for the extremely large overabundances inferred to exist in the atmospheres of Ap stars, which S12 found so difficult to believe.

The results of the magnetic mapping carried out simultaneously with the reconstruction of the Fe abundance are presented in Fig.~\ref{fig:map_fld}. It is reassuring to see very little difference between the \inva\ and \invb\ magnetic maps, despite a more rigourous treatment of the stellar atmosphere by the latter code. Even the difference map (Fig.~\ref{fig:map_fld}c) hardly shows any significant structures. The mean difference is about 0.05--0.10~kG for all three magnetic field vector components, while the standard deviation is 0.19, 0.24 and 0.13~kG for the radial, meridional and azimuthal field, respectively. This corresponds to a typical local magnetic field error of only 2--3\% of the full field amplitude, while even the largest difference of $\sim$\,0.5~kG yields a 7\% local field strength error. The fine structure of the stellar magnetic field, initially found using \inva\ and most clearly seen in the rectangular map of the azimuthal field component, is also faithfully reproduced by \invb. 

This excellent agreement between the magnetic maps recovered using different approaches to the atmospheric structure proves that MDI based on high-quality linear and circular polarisation data is remarkably robust and is insensitive to the possible errors in interpretation of the Stokes $I$ spectra related to using a single model atmosphere. Our results convincingly demonstrate that the bleak S12 assessment of this magnetic inversion procedure is grossly in error. It appears that their allegations are based on an unreasonable extrapolation of a very limited set of forward Stokes parameter calculations for an artificial abundance distribution not resembling any real Ap star. The outcome of the inversions proper, presented here, proves that the methodology used by S12 is misleading and erroneous. A general conclusion, which inevitably follows from this discussion, is that the S12 assertions are largely irrelevant in the context of modern MDI studies by KW10, \citet{kochukhov:2002b,kochukhov:2004d}, and \citet{luftinger:2010}.

\section{Conclusions and discussion}
\label{discussion}

\subsection{Summary of main results}

In this paper we have confronted the criticism by \citet[][S12]{stift:2011} of the basic methodological foundations and specific results of magnetic and abundance Doppler imaging (MDI, DI) of Ap stars. We demonstrated that the doubts about general issues, such as stability of magnetic inversions, robustness of the regularisation procedure, and detectability of small-scale stellar surface features are unwarranted because they were adequately addressed by a number of previous publications ignored by S12. Many successful applications of MDI to slowly rotating stars and the very possibility of photometric mapping of star spots show that the theoretical estimate by S12 of the surface resolution attainable by stellar surface imaging is wrong.

For the specific case of the MDI analysis of \cvn\ by KW10, we showed that S12 have greatly overestimated the contribution of the surface zones with a high Fe abundance to the disk-integrated Stokes spectra and have thus performed their model calculations for an unrealistic parameter range. Using a state-of-the-art opacity sampling model atmosphere code and considering representative Fe abundances, we demonstrated that the effect of metallicity dependence of the model atmosphere structure is negligible for all local Stokes profiles except those corresponding to the strongest lines formed in the most Fe-rich regions ($\log(N_{\rm Fe}/N_{\rm tot})\ge-2$). At the same time, the flux redistribution characteristic of the Fe-rich models leads to a significant continuum brightening of the zones covered by overabundance spots -- an effect potentially important for the spectrum synthesis of Ap stars and for Doppler inversions.

Using a new magnetic DI code which fully incorporates appropriate local model atmospheres in polarised spectrum synthesis, we performed forward calculations of the four Stokes parameter profiles of the \fe\ 5018.44~\AA\ line in the spectrum of \cvn. This experiment showed that using individual model atmospheres for overabundance regions compared to the usual single model atmosphere assumption yields a small modification of Stokes $I$ but is completely negligible for polarisation profiles. In particular, the resulting changes of the Stokes $Q$ and $U$ spectra of \cvn\ do not exceed the observational noise of the data used by KW10 and are significantly smaller than the signature of small-scale magnetic field spots.

Taking advantage of the capabilities of the new magnetic DI code, we have performed the first simultaneous magnetic and Fe abundance inversion using a grid of model atmospheres appropriate for the regions with different elemental abundance. The resulting revised map of the Fe distribution over the surface of \cvn\ changes slightly with respect to the results published by KW10 but maintains the same overall spot topology and still shows the very high Fe overabundances objected to by S12. Our new inversions convincingly demonstrate that, although possibly an artefact of using very strong lines, the most extreme Fe concentrations are definitely not a result of adopting a single mean model atmosphere.

At the same time, the magnetic map produced using the new inversion method turned out to be practically identical to the one inferred by KW10 using a less sophisticated approach. In particular, the presence of small-scale magnetic spots on the surface of \cvn\ is confirmed. The new inversions demonstrate that reconstruction of the magnetic field configuration is sufficiently decoupled from the abundance mapping to be insensitive to the details of model atmosphere structure. These results strongly suggest that self-consistent MDI incorporating model atmospheres calculated for individual local elemental abundances is an unnecessary luxury, not justified by the quality of currently available full Stokes parameter data.

\subsection{Comparison with ZDI of cool stars}

It is instructive to compare the impact of using a single model atmosphere in the MDI of Ap stars with the repercussions of usual assumptions of cool-star ZDI. Although not mentioned explicitly by S12, their criticism obviously applies to the magnetic mapping studies of cool active stars as well. With a few recent exceptions \citep{carroll:2007,kochukhov:2009d}, these studies interpret circular polarisation completely neglecting temperature or brightness inhomogeneities. They also adopt a simplified, temperature-independent analytical descriptions of the local Stokes $I$ and $V$ line profiles instead of calculations based on realistic model atmospheres for different effective temperatures. In contrast, before the present study, MDI analyses of Ap stars have used model atmospheres to calculate the local Stokes profiles as a function of chemical abundance but have ignored continuum intensity variations.

How important is the continuum brightness effect due to chemical inhomogeneities compared to that of cool spots on the surfaces of solar-type stars? Using the {\sc Marcs} model atmosphere grid \citep{gustafsson:2008}, we determined that a 40\% continuum intensity contrast (inferred for a rather extreme case of the Fe spots on \cvn) corresponds to $\approx$\,600~K temperature difference between a spot and photosphere for stellar parameters close to solar values. This is a moderate temperature contrast relative to $T_{\rm phot}-T_{\rm spot}=1500$--2000~K inferred for early-G active stars \citep{berdyugina:2005}. This comparison shows that the errors due to neglecting continuum intensity variation in the magnetic DI of early-type stars are significantly smaller than in the ZDI of cool active stars.

\subsection{DI maps and photometric variability of Ap stars}

DI/MDI techniques have been successfully applied to explain rotational variability of line profiles in various CP stars. The resulting abundance maps allowed to explore the distribution of different chemical elements over stellar surfaces and compare these distributions with the predictions of atomic diffusion theory \citep*{michaud:1981}. On the other hand, for many decades magnetic CP stars were known to exhibit photometric variations with the same period as the magnetic and spectroscopic changes, but this variability had been never explained quantitatively. The fact that the photometric variability occurs in anti-phase in the far-UV and optical spectral regions \citep{molnar:1973,molnar:1975} suggested that the global flux redistribution caused by the phase-dependent line absorption related to chemical spots is the most likely cause of the observed photometric variability. But it was only recently that this hypothesis was rigourously tested by \citet{krticka:2007}, who presented a successful attempt to interpret $uvby$ light curves of the hot Bp star HD\,37776 using DI maps of He and Si derived by \citet{khokhlova:2000}. The adopted photometric variability analysis method was straightforward and realistic: they computed a grid of model atmospheres taking into account the local abundances provided by DI maps and then performed surface integration of the local continuum intensities to obtain phase-dependent fluxes in wavelength regions covered by different photometric filters. This local model atmosphere analysis allowed \citet{krticka:2007} to reproduce the amplitudes of the observed light variability in different filters and even to match the light curve shapes (which evidently depend on the geometrical distribution of the bright and dark areas on stellar surface).
 
This work has provided an important and \textit{completely independent} confirmation of the results of DI by \citet{khokhlova:2000}. Subsequently, the same procedure of explaining the observed light variability by an inhomogeneous surface distribution of chemical elements was successfully applied to several other magnetic CP stars: HR\,7224 \citep{krticka:2009}, $\varepsilon$\,UMa \citep{shulyak:2010b}, and CU\,Vir \citep{krticka:2012}. In all these investigations, the \textsc{Invers} family of Doppler imaging codes was used to derive abundance distributions assuming a single model atmosphere in mapping. Furthermore, using the same chemical spot imaging methodology, \citet{luftinger:2010a} have demonstrated a perfect coincidence between the location of spots of iron-group elements on the surface of the Ap star HD\,50773 and the distribution of bright regions inferred from the high-precision CoRoT light curve of this star.

These results obtained from light curve modelling studies definitely proved that an inhomogeneous distribution of chemical elements is the dominant source of the observed photometric variability and that the usual DI inversions correctly predict the location and amplitude of the main abundance features which give rise to photometric variability. Obviously, a major geometrical artefact in the spot distribution or a gross overestimate of the Fe-peak abundances, as suspected by S12, would not allow reproduction of the stellar light curves. It is quite surprising that this entire series of publications, providing a convincing independent validation of the abundance DI results, has escaped the attention of S12.

\subsection{Final comments}

Besides addressing specific scientific concerns voiced by S12 using the new Stokes profile calculations and drawing readers' attention to copious scientific literature supporting our conclusions, we are led to comment on a number of cases where the S12 interpretation of published studies is subjective and could potentially be misleading.

The choice by S12 of the abundance DI studies yielding ``unrealistic'' elemental overabundances is highly selective, confined to two extreme results, both published in conference proceedings \citep{kuschnig:1998a,piskunov:1998}. This ignores the majority of DI studies published in the refereed literature which reported much more modest abundance contrasts.

S12 compare the MDI studies of \cvn\ by \citet{kochukhov:2002b} and KW10 at great length. Excessive emphasis is given to the most extreme values of the magnetic and abundance maps; however, the maps' information content extends far beyond these extrema.
In fact, the surface distributions agree reasonably well and any differences are well understood. KW10 presented two magnetic maps: one obtained from the Stokes $IV$ spectra and another derived from all four Stokes parameters. There is no major difference between the former and the results of circular polarisation modelling by \citet{kochukhov:2002b}, except for a complementary choice of the inclination angle by KW10 (Stokes $IV$ data do not allow distinguishing $i$ and $180\degr-i$, see, e.g., \citet{bagnulo:2000}) and a systematically weaker field obtained in the more recent study. The latter effect is probably due to an excessive longitudinal field curve regularisation applied by \citet{kochukhov:2002b}. The overall global field topology, characterised by the distribution of the radial field component, is very similar in all magnetic maps reconstructed for \cvn. On the other hand, the discrepancy between the two and four Stokes parameter inversions is the central point of the study by KW10. This difference indicates a rich information content of linear polarisation, not an inversion artefact.

The Fe and Cr maps presented by \citet{kochukhov:2002b} and KW10 agree qualitatively. All maps show a relative underabundance around phase 0, with spots located on both sides of the stellar disk yielding profile doubling. \citet{kochukhov:2002b} used higher resolution data and weaker lines, so their maps are more reliable. The limitations of the abundance mapping using very strong metal lines (weakly sensitive to abundance variation and likely affected by vertical stratification) have been explicitly acknowledged by KW10 (Section 4.2): \textit{``we do not expect to achieve the same quality of abundance inversions as demonstrated by Kochukhov et al. (2002)''}. 

Finally, S12 state that the large overabundances inferred from DI are at odds with the predictions of theoretical diffusion calculations, which have not been shown to produce such large accumulations. In fact, both cited theoretical diffusion studies have artificially limited the maximum accumulation of chemical elements to $\log(N_{\rm el}/N_{\rm tot})=-2.3$ \citep{leblanc:2009} or $\log(N_{\rm el}/N_{\rm tot})=-3.0$ \citep{alecian:2007} in order to avoid numerical instabilities. However, in both studies this limit has been reached for certain chemical elements at some optical depths (see Fig.~3 of \citet{alecian:2007} and Fig.~4 of \citet{leblanc:2009}). Thus, the lack of very high elemental overabundances in the equilibrium diffusion calculations is an artefact of the limited computational capabilities of the currently used atomic diffusion codes.

\section*{Acknowledgments}
OK is a Royal Swedish Academy of Sciences Research Fellow, supported by grants from Knut and Alice Wallenberg Foundation and Swedish Research Council. GAW acknowledges Discovery Grant support from the Natural Sciences and Engineering Research Council of Canada (NSERC). DS acknowledges Deutsche Forschungsgemeinschaft (DFG) Research Grant RE1664/7-1.

\end{document}